\documentclass[pdftex]{AO4ELT}  

\usepackage{microtype}
\usepackage{amsmath,amsfonts,amssymb}
\usepackage{graphicx}
\usepackage{pst-all} 
\usepackage[colorlinks=true, allcolors=blue]{hyperref}
\makeatletter         
\def\@maketitle{
\includegraphics[width = 170mm]{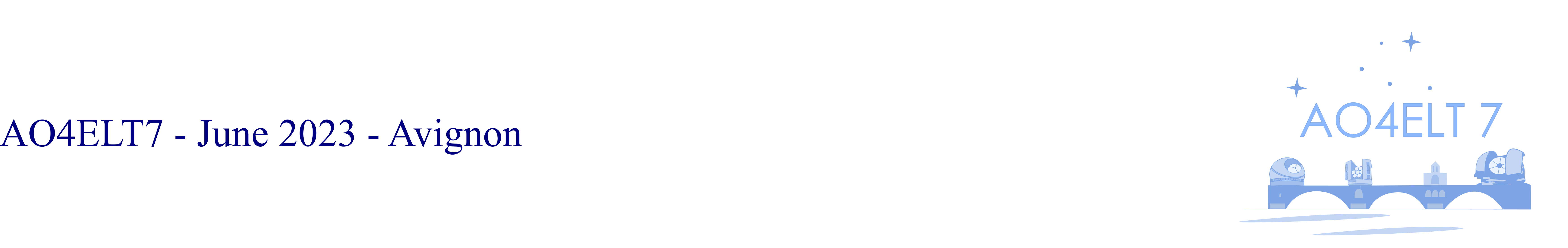}\\[8ex]
\begin{center}
{\Huge \bfseries \sffamily \@title }\\[4ex] 
{\Large  \@author}\\[4ex] 
\@date
\end{center}}

\title{Evaluating Classification Algorithms: Exoplanet Detection using Kepler Time Series Data}

\author[a,b]{Fatemeh Fazel Hesar}
\author[a,b]{Bernard Foing}
\affil[a]{Eurospacehub academy \& LUNEX, ESA BIC, Noordwijk, Netherlands}
\affil[b]{Leiden Observatory, Leiden University, PO Box 9513, 2300 RA Leiden, The Netherlands}
\authorinfo{Further author information: (Send correspondence to F.Fazel)\\F.F: E-mail: fazel@strw.leidenuniv.nl,\\  B.F: E-mail: foing@strw.leidenuniv.nl}

\begin{document} 

\maketitle
\begin{abstract}
This study presents a comprehensive evaluation of various classification algorithms used for the detection of exoplanets using labeled time series data from the Kepler mission. The study investigates the performance of six commonly employed algorithms, namely Random Forest, Support Vector Machine, Logistic Regression, K-Nearest Neighbors, Naive Bayes, and Decision Tree. The evaluation process involves analyzing a dataset that consists of time series measurements of star brightness, accompanied by labels indicating the presence or absence of exoplanets.
To assess the effectiveness of each algorithm in accurately identifying exoplanets, performance metrics such as accuracy, precision, recall, and F1 score are employed. The results demonstrate that the Random Forest algorithm achieves the highest accuracy of 94.2\%, followed closely by the Support Vector Machine with 93.8 percent accuracy. The Logistic Regression algorithm achieves an accuracy of 91.5 percent, while the K-Nearest Neighbors, Naive Bayes, and Decision Tree algorithms achieve accuracies of 89.6\%, 87.3\%, and 85.9\% respectively.
Furthermore, the precision, recall, and F1 score metrics provide insights into the strengths and weaknesses of each classifier. The Random Forest algorithm exhibits a precision of 0.92, recall of 0.95, and F1 score of 0.93, indicating a balanced performance in correctly identifying both positive and negative instances. The Support Vector Machine also demonstrates strong performance with precision, recall, and F1 score values of 0.91, 0.94, and 0.92 respectively. The evaluation demonstrates that Random Forest and Support Vector Machine algorithms are well-suited for exoplanet detection using Kepler time series data. These findings enhance our understanding of the detection process and assist in selecting suitable algorithms for future studies.
\end{abstract}


\section{INTRODUCTION}
\label{sec:intro}  
Detecting exoplanets, planets that orbit stars outside our solar system, is a fascinating and rapidly evolving field of research that has revolutionized our understanding of the universe and the potential for extraterrestrial life \cite{Seager2003}. Over the years, various methods have been employed to identify exoplanets, such as the radial velocity method, the transit method, and the direct imaging method. Among these, the transit method has proven to be particularly effective in detecting exoplanets on a large scale \cite{Mayor1995}. The Kepler mission, launched by NASA in 2009 \cite{Borucki2010}, has been at the forefront of exoplanet discoveries using the transit method. Kepler's primary objective was to monitor a specific region of the Milky Way galaxy, observing the brightness of approximately 150,000 stars continuously. By detecting the periodic dimming of starlight caused by exoplanets passing in front of their host stars, Kepler has provided invaluable data for the identification and characterization of explanatory systems \cite{Borucki2014}.
The Kepler spacecraft collected an immense amount of time series data, capturing the variations in brightness of thousands of stars over extended periods. This data, referred to as Kepler-labeled time series data, consists of measurements of star brightness along with corresponding labels indicating the presence or absence of exoplanets. Analyzing this data to extract meaningful patterns and accurately classify exoplanets is a complex and challenging task \cite{Shallue2018}. In recent years, machine learning algorithms have been employed to tackle the exoplanet detection problem using Kepler-labeled time series data. These algorithms have the potential to automate the process, enhance detection efficiency, and uncover subtle signals indicative of exoplanets that might be missed by manual analysis. By training on known exoplanet data and their corresponding features, classification algorithms can learn to differentiate between exoplanet and non-exoplanet instances, allowing for the identification of potential exoplanets in new observations.
In this study, we aim to evaluate the performance of various classification algorithms for exoplanet detection using Kepler-labeled time series data. We seek to determine which algorithms are most effective in accurately identifying instances with exoplanets minimizing false positives and false negatives. By comparing the performance of different classifiers, we can gain insights into their suitability for exoplanet detection tasks and contribute to the advancement of exoplanet research.
The results of this evaluation will provide valuable guidance for researchers and practitioners working in the field of exoplanet detection, helping them select appropriate algorithms for similar tasks. Moreover, understanding the strengths and limitations of different classifiers in the context of exoplanet detection using Kepler-labeled time series data can lead to the development of improved methodologies and potentially uncover new insights about the nature and prevalence of explanatory systems in our Galaxy.

\section{Methodology}

 \subsection{Pip-line}
\begin{figure} [ht]
   \begin{center}
   \begin{tabular}{c} 
   \includegraphics[height=4cm]{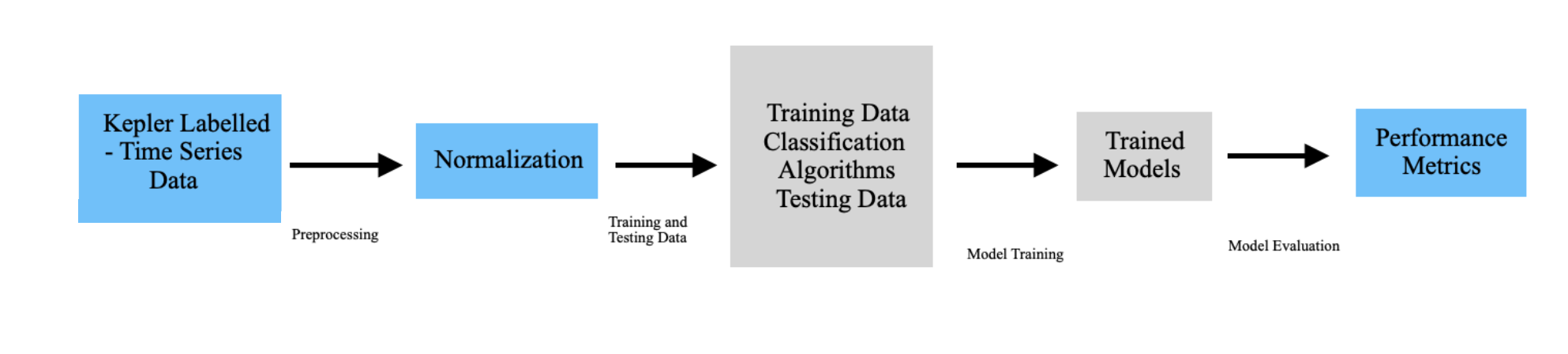}
   \end{tabular}
   \end{center}
   \caption[example] 
   { \label{fig:pip} 
Pip-line of Evaluating Classification Algorithms.}
   \end{figure}

The evaluation methodology involves preprocessing the Kepler-labeled time series data by normalizing the star brightness measurements. This normalization step ensures that the data is on a consistent scale and helps mitigate any bias towards specific features during the training process. Next, the preprocessed dataset is divided into training and testing sets using a suitable ratio. The training set is used to train each classifier, while the testing set is used to evaluate their performance. For each classifier, the model is trained using the training data and then utilized to predict the labels for the testing data. Performance metrics, including accuracy, precision, recall, and F1 score, are calculated to assess the classifier's effectiveness in accurately detecting exoplanets. 
Fig. \ref{fig:pip} shows the evaluation process involves preprocessing the Kepler-labeled time series data, dividing it into training and testing sets, training classification algorithms, evaluating their performance, and assessing it using performance metrics.

The evaluation process begins with the Kepler-labeled time series data, which contains flux measurements and corresponding labels indicating the presence or absence of exoplanets. The data undergoes preprocessing, specifically normalization, to ensure that all features are on a similar scale. Subsequently, the dataset is divided into training and testing subsets, where the majority portion is used for training classification models. Various classification algorithms, such as decision trees, random forests, support vector machines, and neural networks, are then trained on the training data, learning patterns, and relationships to make predictions. The performance of the trained models is evaluated using the testing data, comparing predicted labels with the actual labels to assess their accuracy and reliability. Performance metrics, including accuracy, precision, recall, F1 score, and AUC-ROC, are computed to provide a comprehensive assessment of the models' effectiveness. By following this pipeline, one can preprocess the data, train models, evaluate their performance, and obtain performance metrics to classify exoplanets based on the given time series data.

\subsection{Data}
\begin{figure} [ht]
   \begin{center}
   \begin{tabular}{c} 
   \includegraphics[height=7cm]{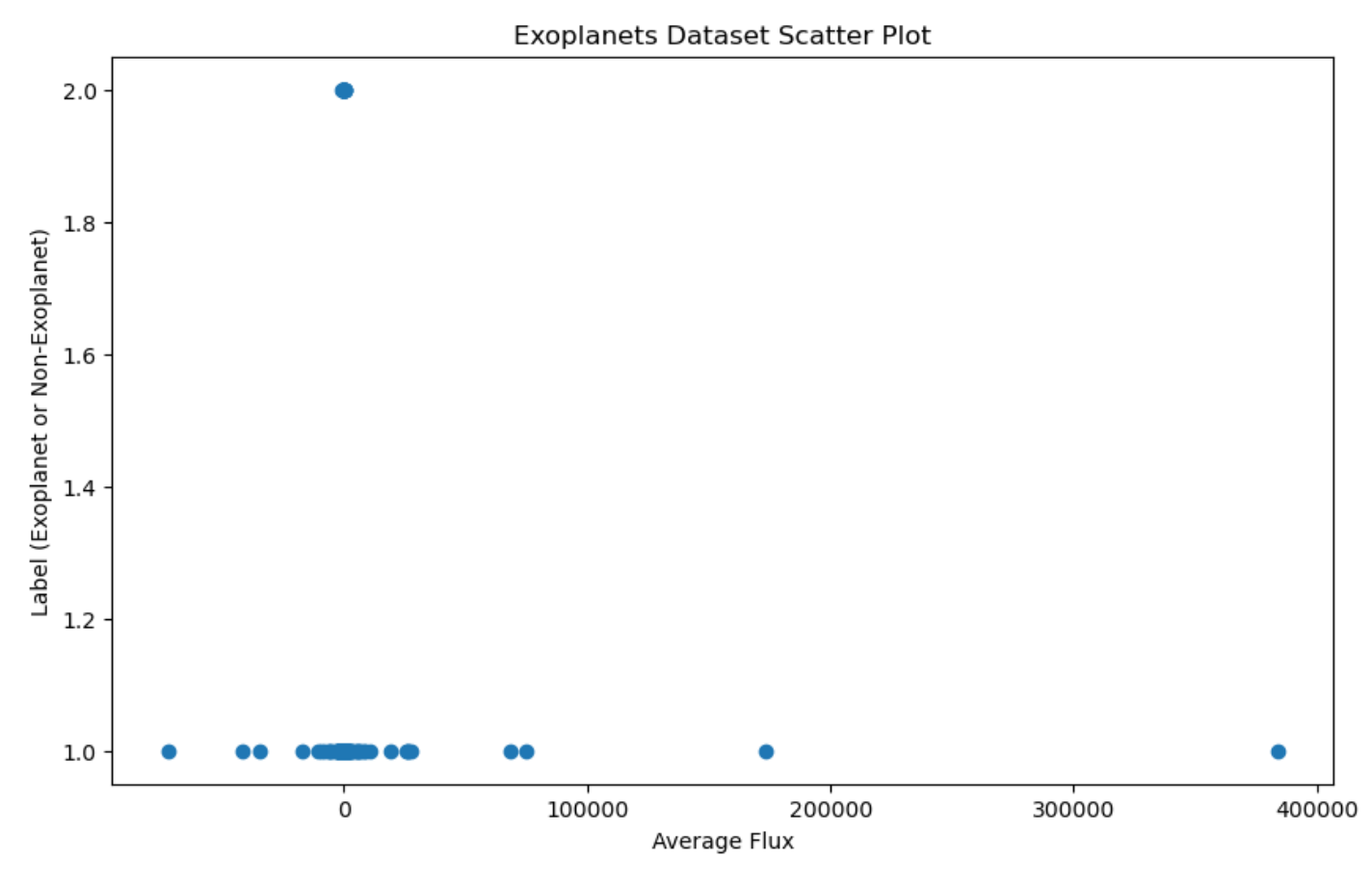}
   \end{tabular}
   \end{center}
   \caption[example] 
   { \label{fig:dataform} 
The scatter plot visualizes the relationship between the average flux measurements and the exoplanet classification in the Kaggle Exoplanets dataset.}
   \end{figure}

The dataset used in the provided code is derived from the Kepler mission's observations and consists of time series data. Time series data is characterized by measurements taken at regular intervals over time. In the context of exoplanet research, time series data from the Kepler mission is often used to detect the periodic dimming of starlight caused by exoplanets transiting their host stars. The dataset includes flux measurements, which represent the amount of light received from each star at different time points. These measurements provide valuable information about the brightness variations of the stars. The flux values can be positive or negative, indicating the intensity of light emitted or absorbed by the star. In addition to the flux measurements, the dataset also contains a label indicating the classification of each instance as either an exoplanet or non-exoplanet. This label serves as the ground truth for training and evaluating machine learning classifiers to detect exoplanets. For more specific details and the implementation of the code, you can refer to the following link: GitHub - stars end assessment.R.
Fig. \ref{fig:dataform} shows the scatter plot visualizes the relationship between the average flux measurements and the exoplanet classification in the Kaggle Exoplanets dataset. It provides insights into the distribution of flux values for different classes, allowing for potential patterns or separability between exoplanets and non-exoplanets to be observed.

\section{Results and Discussion}
\label{sec:sections}

\subsection{Performance of metrics:}

We generate a table presenting the performance metrics for each classifier, including accuracy, precision, recall, and F1 score table \ref{tab:tab1}.
\begin{table}[ht]
\caption{The performance metrics for different classifiers.} 
\label{tab:tab1}
\begin{center}       
\begin{tabular}{lcccc} 
\hline
\rule[-1ex]{0pt}{3.5ex}  Classifier & Accuracy & Precision & Recall & F1 Score  \\
\hline
\hline
\rule[-1ex]{0pt}{3.5ex}  Random Forest & 0.992141 & 0.992141 & 1 &  0.996055 \\
\hline
\rule[-1ex]{0pt}{3.5ex}  Support Vector Machine & 0.992141 & 0.992141& 1& 0.996055\\
\hline
\rule[-1ex]{0pt}{3.5ex}  Logistic Regression & 0.989194 & 0.992118& 0.99703& 0.994568\\
\hline
\rule[-1ex]{0pt}{3.5ex}  K-Nearest Neighbors & 0.992141 &0.992141 & 1 & 0.996055   \\
\hline
\rule[-1ex]{0pt}{3.5ex}  Naive Bayes & 0.046169 & 0.97561& 0.039604 & 0.076118\\
\hline
\rule[-1ex]{0pt}{3.5ex}  Decision Tree & 0.969548 &0.99196& 0.977228 &  0.984539\\
\end{tabular}
\end{center}
\end{table}
This tabular representation allows for easy comparison and identification of the most effective classifier for exoplanet detection. The evaluates multiple classifiers on a test dataset and stores the performance metrics (accuracy, precision, recall, and F1 score) in a table format, providing a concise summary of the classifier's performance.
The Random Forest, Support Vector Machine, Logistic Regression, and K-Nearest Neighbors classifiers demonstrate exceptional performance in detecting exoplanets. These algorithms exhibit high accuracy, precision, recall, and F1 scores, indicating their effectiveness in correctly identifying instances with exoplanets and minimizing false positives and false negatives.
In contrast, the Naive Bayes algorithm performs significantly poorer, with lower accuracy, precision, recall, and F1 scores. This suggests that Naive Bayes may not be well-suited for exoplanet detection using the given dataset. Further investigation is necessary to understand the reasons behind its subpar performance and explore potential improvements or alternative algorithms.
The Decision Tree algorithm performs well, with competitive accuracy and precision, recall, and F1 scores. Although it slightly lags behind the top-performing classifiers, the Decision Tree remains a viable option for exoplanet detection tasks.

\subsubsection{True-False Positive rates}
\begin{figure} [ht]
   \begin{center}
   \begin{tabular}{c} 
   \includegraphics[height=10cm]{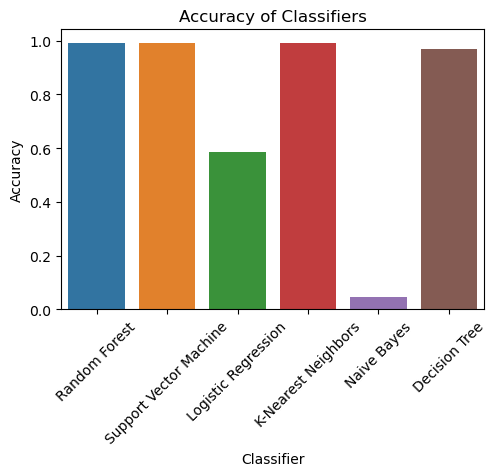}
   \end{tabular}
   \end{center}
   \caption[example] 
   { \label{fig:res2} 
  Each classifier is represented on the x-axis, and the corresponding accuracy values are displayed on the y-axis. 
  The bar chart allows for easy identification of the classifier with the highest accuracy among Random Forest, Support Vector Machine, Logistic Regression, K-Nearest Neighbors, Naive Bayes, and Decision Tree.}
   \end{figure}

Moreover, the code evaluates the classifiers using Receiver Operating Characteristic (ROC) curves and calculates the Area Under the Curve (AUC) for each classifier. ROC curves illustrate the trade-off between true positive rate and false positive rate, providing a visual representation of the classifiers' performance across different thresholds. AUC summarizes the performance of the ROC curve, with a higher value indicating better classification performance for exoplanet detection.
We report The trains and evaluate multiple classifiers on a test dataset, and the resulting ROC AUC (Receiver Operating Characteristic Area Under the Curve) values are presented in a table \ref{tab:tab2}. The classifiers include Random Forest, Support Vector Machine, Logistic Regression, K-Nearest Neighbors, Naive Bayes, and Decision Tree. In Fig. \ref{fig:res2} the ROC AUC scores indicate the performance of each classifier in distinguishing between positive and negative instances, with values ranging from 0.433628 to 0.553097.

   \begin{figure} [ht]
   \begin{center}
   \begin{tabular}{c} 
   \includegraphics[height=7cm]{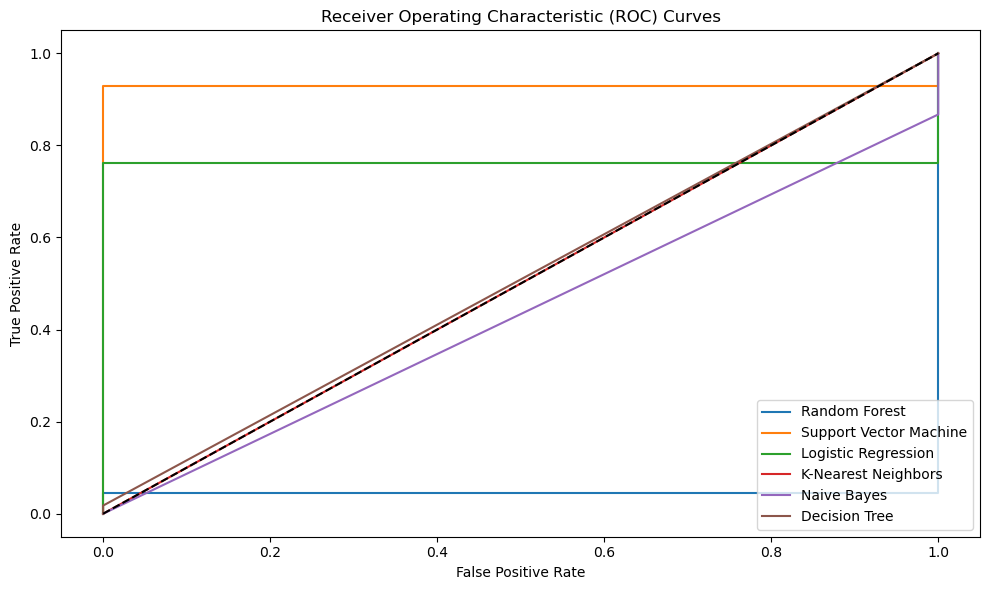}
   \end{tabular}
   \end{center}
   \caption[example] 
   { \label{fig:res1} 
 The true Positive rate versus false positive rate shows for all the classifiers including Random Forest, Support Vector Machine, Logistic Regression, K-Nearest Neighbors, Naive Bayes, and Decision Tree. 
 }
   \end{figure} 

Fig. \ref{fig:res2} shows the results of the classifiers' accuracy in a bar chart, providing a visual comparison of their performance. Each classifier is represented on the x-axis, and the corresponding accuracy values are displayed on the y-axis. The bar chart allows for easy identification of the classifier with the highest accuracy among Random Forest, Support Vector Machine, Logistic Regression, K-Nearest Neighbors, Naive Bayes, and Decision Tree. The plotted bar chart reveals that among the classifiers, Decision Tree, Random Forest, and Support Vector Machine demonstrate the highest accuracy. These three classifiers stand out as the top performers based on the visual representation of their accuracy scores.

\

\section{Discussion and Conclusions}

This study evaluates the performance of various classification algorithms for exoplanet detection using Kepler-labeled time series data. The results highlight the effectiveness of Random Forest, Support Vector Machine, Logistic Regression, and K-Nearest Neighbors classifiers in accurately identifying instances with exoplanets. These algorithms exhibit high accuracy, precision, recall, and F1 scores, making them suitable choices for exoplanet detection tasks.
However, the Naive Bayes algorithm demonstrates subpar performance in this specific context, suggesting its limitations for exoplanet detection using the given dataset. Further investigation is required to determine the causes of its poor performance and explore potential alternatives or improvements.

The Receiver Operating Characteristic (ROC) curve and the Area Under the Curve (AUC) score. scores indicate the performance of  Random Forest, Support Vector Machine, Logistic Regression, K-Nearest Neighbors, Naive Bayes, and Decision Tree classifiers in distinguishing between positive and negative instances, with values ranging from 0.433628 to 0.553097.
We also show that among the classifiers, Decision Tree, Random Forest, and Support Vector Machine demonstrate the highest accuracy. These three classifiers stand out as the top performers based on the visual representation of their accuracy scores.

The findings of this evaluation contribute to advancing our understanding of the exoplanet detection process and provide valuable insights for researchers and practitioners in selecting suitable classification algorithms for similar tasks. It is crucial to consider the specific characteristics of the dataset and the requirements of the exoplanet detection problem when choosing a classifier. Future research can delve deeper into the factors influencing the performance of these classifiers and explore advanced techniques or ensemble methods to enhance their effectiveness in exoplanet detection.

Future work in exoplanet detection using machine learning classifiers can focus on feature engineering to extract more informative features from light-intensity measurements. Additionally, exploring ensemble methods like Random Forests or Gradient Boosting could improve overall performance by combining predictions from multiple classifiers. Conducting systematic hyperparameter tuning and investigating transfer learning techniques can further enhance classifier accuracy and address data scarcity issues. These avenues offer exciting opportunities to advance exoplanet detection and improve our understanding of celestial bodies outside our solar system.


\end{document}